\documentclass[letterpaper, a4paper]{amsart}

\usepackage{amsmath}
\usepackage{latexsym}
\usepackage[all]{xy}
\usepackage{color}
\newtheorem{teorema}{Theorem}[section]

\include{amslatex}

\newcommand{\Diff}[0]{{\textup{Diff}}}

\newtheorem{ejemplo}[teorema]{Example}

\numberwithin{equation}{section}

\begin{document}
\begin{title}[Emergent quantum mechanics and quantum non-local correlations]
 {Emergent quantum mechanics and the origin of quantum non-local correlations}
\end{title}
\clearpage\maketitle
\thispagestyle{empty}
\begin{center}
\author{Ricardo Gallego Torrom\'e\footnote{Email: rigato39@gmail.com}}
\end{center}
\bigskip

\begin{center}
\address{Frankfurt Institute for Advanced Studies\\
Ruth-Moufang-Stra{\ss}e 1, 60438 Frankfurt, Germany}
\end{center}

\begin{abstract}
A geometric interpretation for quantum correlations and entanglement according to a particular framework of emergent quantum mechanics is developed. The mechanism described is based on two ingredients: 1. At an hypothetical sub-quantum level description of physical systems, the dynamics has a regime where it is partially ergodic and 2. A formal projection from a two-dimensional time mathematical formalism of the emergent quantum theory to the usual one-dimensional time formalism of quantum dynamics. Observable consequences of the theory are obtained. Among them we show that quantum correlations must be instantaneous from the point of view of the spacetime description, but the spatial distance up to which they can be observed must be bounded. It is argued how our mechanism avoids Bell theorem and Kochen-Specken theorem. Evidence for non-signaling faster than the speed of light in our proposal is discussed.
\end{abstract}

\section{Introduction}
Non-local quantum correlations are among the most counterintuitive quantum phenomena. Probably the most surprising fact about them is their apparent superluminal  speed, which lower bound has been demonstrated to be several orders of magnitude faster than the speed of light in vacuum \cite{CocciaroFaettiFronzoni, GisinScaraniTittleZbinden, Salart et al., Juan Ying et al.}. The possibility to have quantum correlations over macroscopic distances \cite{Bancal et al} is also rather counterintuitive. If we agree that quantum correlations corresponds to real physical processes, their characteristic urge us to find a rational mechanism for them. There are several grounds for this urgency. The first is that their apparent unlimited speed does not fit with the rest of physical phenomena that it is know. Therefore, an explanation is mandatory.  Second, although quantum mechanics successfully predicts the statistical properties of the physical observables associated with entangled quantum systems,  it does not offer a geometric interpretation of quantum correlations. This lack of  geometric description in the framework of quantum mechanics, a theory whose dynamics has a spacetime interpretation in the form of sum over histories  \cite{Feynman1948} is, from our point of view, an unsatisfactory situation.

Of course, this interpretation is not correct if a positivistic interpretation of quantum phenomena is adopted. However, there is a persistent problem with this posture. The theoretical and conceptual structure of another fundamental pilar of modern theoretical physics, general relativity, is in sharp confrontation with the description that quantum mechanics offers of physical dynamical processes. General relativity is a theory where it is possible to attach for many physical systems, at least partially, an history, that is, an objective, real evolution in spacetime. In general relativity physical processes happen in the spacetime independently of the observation realized by any observer, although the particular description of the processes depends on the particularities of each observer. Furthermore, observers' history happen in the spacetime too. This property of general relativity contrasts with the {\it instrumentalist attitude} towards quantum mechanics, where the theory is regarded as a {\it complete and ultimate description of physical systems}. It could also conflict with other interpretations of quantum mechanics, specially if  the apparent superluminal character of quantum correlations is considered to be  a consequence of real physical phenomena. However, this second general perspective on quantum mechanics admits a higher  flexibility, since  within the instrumentalist interpretation  it is difficult to modify quantum mechanics in a consistent way. Indeed, from the instrumentalist point of view, quantum mechanics is a complete, consistent framework, for instance, as the Copenhagen interpretation promotes \cite{Bohr1935}. If we adopt the second type of attitude,  modifications of quantum mechanics or modifications of general relativity or modifications of both theories can be suggested, in  attempts to find an unifying framework to describe the overall our physical experiences, in particular  the phenomena that involves quantum correlations.

In this paper the second point of view is adopted. A partial  justification for this attitude is the hypothesis that Nature is ruled by a set of harmonic  fundamental laws such that they must either explain the complete set of phenomenological experiences that we acknowledge or justify new emergent laws at certain regimes of dynamics. This set of experiences includes the phenomena described by general relativity and the acceptance of how   general relativity explain them: for the phenomena of gravitational interaction and macroscopic interactions, in a general realistic way in the framework of a dynamical and objectively real spacetime arena. The existence of such macroscopic objective spacetimes imposes several obstructions in the construction of universal schemes. 

This acceptance of the existence of a dynamical objective spacetime does not obviates  the problem of the singularities of the solutions of Einstein equations. Thus there is a need for a modification of general relativity. But by the above argument, we suggest the possibility to use  {\it classical spacetimes} models, that is, models without the possibility of quantum superposition of spacetime structures \cite{Ricardo2017}. In this way, the transition from a fundamental description towards a macroscopic, objective and real spacetime description can be done easier. This theoretical possibility of classical includes the use of discrete models of quantum spacetime, for instance as in causal sets schemes \cite{Dowker Henson Sorkin 2003} or as in Snyder's quantum spacetime \cite{Snyder}. 

Among the alternatives to standard quantum theory, there is a class of theories proposing emergent interpretations of quantum mechanics. The aim of such theories is to recover quantum mechanics or relativistic quantum field models in an appropriate approximation as an effective  description from a deeper level of physical description. An example of emergent quantum mechanics is Bohm's proposal to explain the de Broglie-Bohm theory from a sub-quantum level of physical reality (see for instance \cite{Bohm1952} and \cite{Bohm1982}, {\it chapter} 4). More recent investigations of  emergent theories of quantum mechanics can be found  in  \cite{Adler, Elze, FernandezIsidroVazquez2016, Groessing2013, Ricardo2017, Hooft2016, Smolin2012}, for instance. The particular theory that we will consider in this paper was sketched in \cite{Ricardo06} and further developed in \cite{Ricardo2017} by the author. In such theory it was assumed the existence of a sub-quantum level of physical reality where the degrees of freedom are deterministic and local in an extended configuration space. The theory was called {\it Hamilton-Randers theory} because the Hamiltonian functions are constructed from an underlying {\it Randers-type metric structure} \cite{Randers} defined in the configuration space associated to the fundamental degrees of freedom of the theory, not the quantum degrees of freedom that appear in a classical or quantum description. The postulated fundamental scale is larger than any mass scale associated to the standard model of particle physics. It is unspecified by the theory except by the assumption that the ratio of such scale and any of the scales where quantum mechanics  gives an appropriate description of the dynamical system is large enough to allow the effective application of the mathematical theory of {\it concentration of measure} \cite{Gromov, MilmanSchechtman2001} as it was discussed in \cite{Ricardo2017, Ricardo2015}. Informally, concentration of measure is the mathematical fact that for metric spaces endorsed with a probability measure, $1$-Lipschitz regular functions are almost constant almost everywhere. Concentration of measure is an essential tool in our argument for  the existence of equilibrium regimes of the underlying sub-quantum dynamics and in the  proof of other properties of the dynamics at the sub-quantum level (\cite{Ricardo2017}, {\it sections} 6 and 7). Indeed, one can argue in the framework of Hamilton-Randers theory using in an essential way concentration of measure arguments  that the gravitational interaction and the collapse of the quantum state are two aspects of the same physical phenomena.

 In Hamilton-Randers models the fundamental degrees of freedom are described by  points of a classical configuration space. They obey a {\it fast dynamics} consisting of sequences of {\it fundamental cycles} such that each fundamental cycle is composed by a sequence of consecutive ergodic/contractive/expansive dynamical processes.

The concept of time appears in Hamilton-Randers theory in the form of two different notions. First, as a class of $t$-time parameters describing the evolution of the {\it fast dynamics} of the fundamental cycles. Second, as a class of discrete $\tau$-time parameters associated with the  number of fundamental cycles of a given quantum system or with the emergent character of regular, stable quantum processes. Note that although $\tau$-time parameters are discrete by construction, for practical purposes we can consider them as continuous parameters.
These two types of time parameters are independent from each other. This fact is the origin of the notion of {\it two-dimensional time and double-dynamics}  in Hamilton-Randers theory \cite{Ricardo2017}. This is similar to the notion of {\it fast/slow} dynamical models in classical mechanics, from where we adopt occasionally the terminology. However, since any $\tau$-time parameter is emergent, it cannot be used in the description of the sub-quantum processes from which it emerges. 

The usual perception of {\it change} or {\it pass of time} can be described by using one-dimensional, real time parameters. It is in this sense that one can say that the time used in the description of macroscopic and  quantum mechanical dynamics is one-dimensional. Therefore, there is  a formal projection from a mathematical description of Hamilton-Randers theory with two-dimensional time parameters to the quantum mechanical description with a one-dimensional time parameter. According to our theory, this formal projection is an essential idea for our explanation of the apparent instantaneous spacetime  character of quantum correlations.

 Another relevant aspect of Hamilton-Randers theory is the assumption about certain degree of ergodicity in a sub-domain of the fundamental dynamics. Such ergodicity together with the above mentioned formal projection can potentially explain quantum entanglement in terms of a realistic picture \cite{Ricardo2017}. However, ergodicity alone is not enough to explain the specific features of the quantum correlations, specially their apparently un-bounded value of the speed in the correlations.

The aim of the present paper is to provide a self-contained presentation of a mechanism for quantum correlations according to the theory develop in \cite{Ricardo2017}. The mechanism is based upon general assumptions on the dynamics of the models for sub-quantum physics and their generic mathematical features, namely, the high dimensionality of the configuration spaces and the regularity properties of the dynamics.
We have discussed several consequences of the mechanism and, despite such consequences are of general character, they offer the opportunity to test the theory against experiment. Among the consequences of our theory we remark the necessary instantaneous spacetime character of the quantum correlations and their spacetime range limitation. Also, it is discussed an universal type of process leading to the loose of entanglement. Related with these issues, we briefly discuss how our theory avoids the direct application of Bell inequalities \cite{Bell1964, Bell} and of Kochen-Specken theorem \cite{Kochen-Specken1967}. Moreover, we provide evidence for the absence of signaling faster than the speed of light in Hamilton-Randers theory.

\section{Hamilton-Randers theory}In order to introduce our interpretation of quantum correlations, let us first briefly consider some fundamental aspects of the theory developed  in \cite{Ricardo2017}.
Hamilton-Randers systems are deterministic and local dynamical models for physical processes happening at a hypothesized fundamental scale. They are also relativistic models, in the sense that the coordinate speeds of the fundamental degrees of freedom are uniformly bounded in some specific local coordinate systems. It is assumed that such upper limit is the speed of light in vacuum.

The configuration space in a Hamilton-Randers model is a product of tensor manifolds of the form
\begin{align}
TM=\,\prod^{N}_{k=1}\,TM^k_4,
\end{align}
where $N\in \mathbb{N}$ is the number of sub-quantum molecules describing a particular quantum system. $N$  is for practical purposes a large number compared with $1\in\,\mathbb{N}$.  $\{M^k_4\}^N_{k=1}$ is a collection of four-manifolds, each of them diffeomorphic to a given {\it model manifold} $M_4$ by means of the collection of diffeomorphisms $\{\vartheta_k:M^k_4\to M_4,\,k=1,...N\}$. On the other hand, $M_4$ is identified with the spacetime manifold where observable events can potentially be localized. The sub-quantum  degrees of freedom are described by points $u=(x,y)\in\,TM$, with $x\in M$. It is assumed the existence of a deterministic, local dynamical law in $T^*TM$ described by an operator $U_t$ that happens at the fundamental scale. The $U_t$ dynamics is discrete. However, it is convenient to consider the continuous approximation. In the continuous version of the theory the $U_t$ dynamics can be associated with a continuous  flow in $T^*TM$. Although the details of the $U_t$ dynamics still need to be clarified further, the specific form is not necessary for the purpose of this paper. There is also defined on the co-tangent space $T^*TM$ a probability measure $\mu_P$ and a {\it quasimetric} of dual Randers type, making the co-tangent space $T^*TM$ a measure-metric space \cite{Gromov}.

 One of the fundamental assumptions of the theory proposed in \cite{Ricardo2017} is that the number of sub-quantum degrees of freedom of Hamilton-Randers systems is in proportion $N$ to $1$ with the quantum degrees of freedom of the system. For example,  the dynamical description of an electron corresponds in Hamilton-Randers theory to a dynamical system with a large number $N_e\gg 4$ of degrees of freedom whose dynamical variables are points of $T^*TM$. This assumption leads to the idea that quantum particles are complex dynamical systems. The parameter $N$ varies depending on the particular quantum system.

It is also postulated that when the system is not interacting with the environment, the $U_t$ dynamics is perfectly cyclical. Each of the {\it fundamental cycles} is characterized by a semi-period $T$, which is given by the expression
\begin{align}
 \log\left(\frac{T}{T_{min}}\right)=\,\frac{T_{min}\,Mc^2}{\hbar},
\label{ETrelation}
\end{align}
where $M$ is the mass of the quantum system and $T_{min}$ is the minimal semi-period allowed for any physical system. The semi-period $T_{min}$ is a constant that depends on the choice of the units of the arbitrary $t$-parameter but that otherwise is the same for all Hamilton-Randers systems. The relation \eqref{ETrelation} can be justified by the fact that it implies the additivity of the mass parameter $M$ for non-interacting quantum systems under the corresponding multiplicative property of the semi-period $T$ and because it can be used to justify the {\it energy-time quantum uncertainty relation} from an underlying realistic framework (\cite{Ricardo2017}, {\it section} 3).

A main technical tool in Hamilton-Randers systems is the theory of concentration of measure \cite{Gromov,MilmanSchechtman2001}, which is present in Hamilton-Randers models  due to the high dimensionality of the configuration space $TM$. What concentration does is to ensure Gaussian deviation estimates from the Levy's mean value in the case when the functions are regular enough ($1$-Lipschitz functions). The deviation from the Levi's mean that these functions can have is very small almost everywhere for large dimensional spaces, as it is the case of Hamilton-Randers systems. Concentration of measure is applied in Hamilton-Randers theory under the assumption that in each fundamental cycle there is a {\it contractive regime} where the $U_t$ dynamics is $1$-Lipschitz in the sense of an operator norm. In contrast with {\it objective collapse models} \cite{Diosi,GhirardiGrassiRimini,GhirardiRiminiWeber,Penrose}, the collapse in Hamilton-Randers theory happens naturally in any quantum system and without the need of interaction with external macroscopic measurement devices. It happens for every quantum system. After each concentration process finishes, there is an expanding process that completes the fundamental cycle. After the expanding domain finishes and if there is no measurement process undertaken by an external observer or external device or if the system do not interact with other particles, a new cycle with a ergodic regime stars and so on... We called these concentration process {\it natural spontaneous collapse}, to emphasize its different mechanism and nature from the processes that appear in the literature under the spell of objective collapse models.

It was also argued that  during the concentration regime the $U_t$ dynamics is dominated by a classical gravitational interaction. This claim was based on several formal analogies between the properties of classical gravitational interaction and the $U_t$ dynamics in the contractive regime. Among these analogue properties are diffeomorphism invariance, relativistic causal cones and the weak equivalence principle. Indeed, using concentration of measure theory we were able to derive from the principles of Hamilton-Randers theory that the weak equivalence principle must hold (\cite{Ricardo2017}, {\it section 7}).

The ergodic regime in each fundamental cycle has some specific characteristics (see \cite{Ricardo2017}, {\it section 5} for details of the characteristics  ascribed to these dynamical domains of the $U_t$ evolution). The first one is that it is not absolutely ergodic. Spacetime is not totally filled by the evolution, only the {\it allowable causal domain} of the spacetime is filled. Similarly, the allowable velocity space is diffeomorphic to a cone in  $T_x M_4$ and is filled ergodically during each fundamental cycle.  These characteristics  imply the use of a modified version of the ergodic hypothesis, namely, the time average operation  is equivalent to the speed coordinate average operation, leaving the spacetime coordinates as labels. These labels indicate where in spacetime experiments can be performed. Despite this dynamics is not strictly ergodic, we will call it ergodic but keep in mind the weaker meaning of this term respect to the standard use in dynamical systems theory.

 Besides the $t$-time parameter in the description of the $U_t$ dynamics, in Hamilton-Randers models there are  $\tau$-time parameters, which can be used as physical time parameters of the effective quantum and classical dynamical systems. As we mentioned before,  the $\tau$-time parameters are associated with the counting of fundamental cycles. Note that these counting of the number of cycles by an arbitrary observer depend on the state of motion or other physical attributes of the observer or physical clock. Therefore, the notion of $\tau$-time parameter does not correspond to any kind of absolute time parameter and indeed the theory of Hamilton-Randers systems is consistent with the principle of relativity  and diffeomorphism invariance \cite{Ricardo2017}. Both parameters $(t,\tau)$ are discrete, for consistence of the theory. However, we are considering the approximation where $(t,\tau)$ are continuous. The justification is given in \cite{Ricardo2017}.

The relevant difference between the $\tau$-time and $t$-time is reflected in the impossibility to use any $\tau$-time parameter (a slow-time parameter) to describe the $U_t$ dynamics. This is because for each lapse of $\tau$-time, a whole fundamental cycle of $U_t$ evolution has been completed. Thus the $U_t$ dynamics is analogous to a {\it fast dynamics} as it appears in classical mechanics and statistical mechanics. The {\it slow dynamics} $U_\tau$ is identified with the usual quantum dynamics by application of Koopman-von Neumann theory of dynamical systems in the approximation where $\tau$-time parameters are continuous.

Let us describe some details of the dynamics of Hamilton-Randers systems. First, let us recall that the $t$-time dependent Hamiltonian function for the $U_\tau$ dynamics is postulated to be of the form \cite{Ricardo2017}
\begin{align}
H(u,p,t)=\,\left(1-\kappa(t)\right)\,\left(\,\sum^{8N}_{k=1}\,\beta^k(u)p_k\right),
\label{Hamiltonianfunction}
\end{align}
where $\kappa$ is a scalar factor determined by the $U(t)$ dynamics. The on-shell constraints
\begin{align}
\dot{x}^i=\,y^i,\quad i=1,2,...,N.
\end{align}
are also imposed.

It is postulated that during each of the fundamental cycles there is a contractive regime such that the $U_t$ dynamics is $1$-Lipschitz continuous in some operator norm \cite{Ricardo2017}. In such regime the conditions
\begin{align}
\lim_{t\to (2n+1)T}\, H=0,\quad \lim_{t\to (2n+1)T}\,\left(1-\kappa(t)\right)=0, \quad n\in\,\mathbb{Z}
\label{equilibriumlimit}
\end{align}
hold good and the dynamics is manifestly $\tau$-time diffeomorphic invariant, since the Hamiltonian function \eqref{Hamiltonianfunction} is zero or close to zero in such regime of the dynamics.

We apply Koopman-von Neumann theory \cite{Koopman1931, Von Neumann} to Hamilton-Randers systems in the following way. First, we introduce the following quantization prescription, which is indeed an statement on the existence of a non-commutative Lie algebra,
\begin{align}
(u,p)\mapsto (\hat{u},\hat{p}),\quad
[\hat{u}^i,\hat{u}^j]=0,\quad [\hat{p}_i,\hat{p}_j]=0,\quad [\hat{u}^i,\hat{p}_j]=\,\delta^i_{j},\quad i,j=1,...,8N.
\label{quantumalgebra}
\end{align}
This quantization prescription is not equivalent to the standard canonical quantization of  quantum mechanics, since in our case half of the $u$-coordinates represent velocity coordinates. After quantization, all the  $\hat{u}$-velocity $\{\hat{y}^i\}^{4N}_{i=1}$ elements of the quantum algebra \eqref{quantumalgebra} commute with the $\hat{u}$-position coordinate operators $\{\hat{x}^i\}^{4N}_{i=1}$. This situation contrasts with the usual canonical quantization used in quantum theory, where the canonical position operators and the velocity operators do not commute. The fact that the dynamics of sub-quantum degrees of freedom is deterministic is not in contradiction with the uncertainty principle in quantum mechanics, since the degrees of freedom represented by points $u\in TM$ are not the same than the quantum degrees of freedom used in the quantum mechanical description of the system.

 Let us consider a linear representation of the algebra \eqref{quantumalgebra} on a vector space $\mathcal{H}_{F}$. A collection $\{|u\rangle\}\subset \mathcal{H}_F$ of eigenstates for the $\hat{u}$ operators is also introduced. Then the  vector space $\mathcal{H}_{F}$ can be furnished with a natural scalar product in such a way that $\mathcal{H}_{F}$ with such scalar product is a pre-Hilbert space \cite{Ricardo2017}. We assume that indeed $\mathcal{H}_F$ is a Hilbert space. Then we can state that the operators $\{\hat{u}^\mu_k\}$ and $\{p^\mu_k\}$ are self-adjoint to respect the inner product in $\mathcal{H}_F$. Furthermore, the eigenvalues of $\{\hat{u}^\mu_k\}$ are continuous and compatible with the atlas structure of $TM$.

 The eigenvectors of $\{\hat{u}^\mu_k\}$ are such that
 \begin{align*}
 \hat{x}^\mu_k\,|x^\mu_k,y^\mu_k\rangle=\,x^\mu\,|x^\mu_k,y^\mu_k\rangle,\quad  \hat{y}^\mu_k\,|x^\mu_k,y^\mu_k\rangle=\,y^\mu\,|x^\mu_k,y^\mu_k\rangle.
 \end{align*}
In terms of the generator system of $\mathcal{H}_F$
\begin{align*}
\{|u^\mu\rangle\}^N_{k=1}\equiv\left\{|x^\mu_k,y^\mu_k\rangle,\, N=1,...,N,\, \mu=0,1,2,3\right\}
\end{align*}
a generic element of $\mathcal{H}_F$ is of the form
\begin{align}
\psi(x)=\,\sum^N_{k=1}\,\frac{1}{\sqrt{N}}\,\int_{T_xM^k_4}\,d^4z_k \,e^{\imath \varphi_k(x_k,z_k)}\,n_k(x^\mu_k,z^\mu_k)\,|x_k,z_k\rangle,
\label{quantumstates}
\end{align}
where in the following $|x_k,z_k\rangle$ stands for $|x^\mu_k,z^\mu_k\rangle$. $\vartheta_k:\mathcal{M}^k_4\to M_4$ are diffeomorphisms and we denote by $\vartheta^{-1}_k(x)=x_k$, etc...
 Note that the velocity coordinates are integrated. This is an expression of the ergodicity in one of the dynamical regimes of the $U_t$ dynamics. The ergodic theorem is only applied respect to the speed coordinates and not respect to the spacetime coordinates. The spacetime coordinates are labels for the degrees of freedom. The choice of the type of label is irrelevant. Hence, the theory must be invariant under diffeomorphism invariance in $M_4$. This invariant condition is totally consistent with Hamilton-Randers theory.

 The Hamiltonian operator $\widehat{H}$ is obtained from  the Hamiltonian function \eqref{Hamiltonianfunction} by application of Born-Jordan quantization prescription to the algebra \eqref{quantumalgebra}. The quantized Hamiltonian $\widehat{H}$ defines the $U_\tau$ dynamics through the corresponding Heisenberg equation. In the equilibrium domain $\widehat{H}=0$, that can be read in a weak way as $\widehat{H}\psi=0$, where $\psi$ represents a quantum state of the system.

\section{Entanglement from the perspective of Hamilton-Randers theory and a mechanism for quantum correlations}
Let us consider a dynamical system of two quantum  particles $a$ and $b$, not necessarily of the same type. At the level of the sub-quantum degrees of freedom, the dynamical system is described in the Koopman-von Neumann formulation of the dynamics by an element of the Hilbert space $\mathcal{H}_F$ of the form
 \begin{align*}
 \Psi_{a \otimes b}(u_a,u_b) =\,&\Big(\sum^{N_a}_{k=1}\Big( n_{1k}(x_k,z_k)\,\exp(\imath \,\varphi_{1k}(x_k,z_k))\\
 & +n_{2k}(x_k,z_k)\,\exp(\imath \,\varphi_{2k}(x_k,z_k))\Big)|\vartheta^{-1}_{k}(x),z_k\rangle\Big)\otimes\\
 &\Big(\sum^{N_b}_{l=1}\,\Big(n_{1l}(x_l,z_l)\,\exp(\imath \varphi_{1l}(x_l,z_l))\\
 & +n_{2l}(x_l,z_l)\,\exp(\imath \,\varphi_{2l}(x_l,z_l)) \Big)|\vartheta^{-1}_{l}(x),z_l\rangle\Big),
 \end{align*}
 with $N_a+N_b=N$.
 In this expression the indices $k$ and $l$  run over different sets of sub-quantum degrees of freedom, each set is associated to the quantum particles $a$ or $b$ respectively. On the other hand, the indices $1$ and $2$ refer to two different subsets of sub-quantum degrees of freedom associated directly with experimental arrangements $1$ and $2$ performed by observers $1$ and $2$ at possibly different spacetime locations.
  These two classifications of the same collection of sub-quantum degrees of freedom do not coincide in general. Thus, there are degrees of freedom associated to $a$ which are associated to $1$ or to $2$ and analogously for $b$. The way in which they are associated is determined by the $U_t$ dynamics and the experimental arrangements. It must be consistent with the quantum mechanical description and Born's rule. These aspects of the theory are discussed in  {\it section 5} and {\it section 6} in \cite{Ricardo2017}.
Note that there are other formal combinations, with complex valued relative phases between the vector states associated to sub-quantum degrees of freedom at $1$ and at $2$. Let us also emphasize  that since the sub-quantum degrees of freedom are classical, the conditions $n_{1k}(x_k,y_k)=0$ or $n_{2k}(x_k,y_k)=0$, etc... must hold good.

 The {\it entanglement conditions} are {\it linear constraints} relating the $N_a+ N_b$ sub-quantum degrees of freedom of the quantum particles $a$ and $b$,
\begin{align}
0 = \left| \begin{array}{cr}
 n_{1k}(x_k,z_k)\,\exp(\imath \,\varphi_{1k}(x_k,z_k) & -\, n_{2l}(x_l,z_l)\,\exp(\imath (\varphi_{2l}(x_l,z_l)))\\
n_{2k}(x_k,z_k)\,\exp(\imath \,\varphi_{2k}(x_k,z_k))& n_{1l}(x_l,z_l)\,\exp(\imath \,\varphi_{1l}(x_l,z_l))  \end{array} \right|.
\label{entanglementcondition2}
\end{align}
The number of these constraints is of order $N_a\,\cdot N_b$, which is a large number compared with $N_a+N_b$, for $N_a, N_b\gg 1$. This shows that no all the constraints \eqref{entanglementcondition2} can be independent but also that they are very stringent, due to the large number of them.
If these constraints hold good, after averaging on the velocity components,  the corresponding quantum state is
\begin{align}
\psi_{ab}(x)=&\Big(\sum^{N_a}_{k=1} \int_{T_xM^k_4}\,d^4z_k \,n_{1k}(x_k,z_k)\,\exp(\imath \,\varphi_{1k}(x_k,z_k))|x_k,z_k\rangle\Big)\break\\
& \nonumber \otimes \Big(\sum^{N_b}_{l=1} \int_{T_xM^k_4}\,d^4z_l\,  n_{2l}(x_l,z_l)\,\exp(\imath \,\varphi_{2l}(x_l,z_l))|x_l,z_l\rangle\Big)\break\\
 &\nonumber +\Big(\sum^{N_a}_{k=1} \int_{T_xM^k_4}\,d^4z_k \,n_{2k}(x_k,z_k)\,\exp(\imath \,\varphi_{2k}((x_k,z_k))|x_k,z_k\rangle\Big)\break\\
& \nonumber \otimes \Big(\sum^{N_b}_{l=1} \int_{T_xM^k_4}\,d^4z_l\, n_{1l}(x_l,z_l)\,\exp(\imath \,\varphi_{1l}(x_l,z_l))|x_l,z_l\rangle\Big).
\label{entangledstate2}
\end{align}
This is an entangled state of the form
 \begin{equation}
 \psi_{ab}(x)=\,\frac{1}{\sqrt{2}}\Big(\psi_{1a}(x)\otimes\,\psi_{2b}(x)+\psi_{2a}(x)\otimes\,\psi_{1b}(x)\Big),
 \label{nonproductstate}
 \end{equation}
where
\begin{align}
&  \psi_{1a}(x):=\,\sum^{N_a}_{k=1} \int_{T_xM^k_4}\,d^4z_k \,n_{1k}(x_k,z_k)\,\exp(\imath \,\varphi_{1k}(x_k,z_k))|x_k,z_k\rangle,\break\\
&\nonumber \psi_{2a}(x):=\,\sum^{N_a}_{k=1} \int_{T_xM^k_4}\,d^4z_k \,n_{2k}(x_k,z_k)\,\exp(\imath \,\varphi_{2k}(x_k,z_k))|x_k,z_k\rangle,\break\\
&\nonumber \psi_{1b}(x):=\,\sum^{N_b}_{l=1} \int_{T_xM^l_4}\,d^4z_l\, \cdot n_{1l}(x_l,z_l)\,\exp(\imath \,\varphi_{1l}(x_l,z_l))|x_l,z_l\rangle,\break \\
&\nonumber \psi_{2b}(x):=\,\sum^{N_b}_{l=1} \int_{T_xM^l_4}\,d^4z_l\, \cdot n_{2l}(x_l,z_l)\,\exp(\imath \,\varphi_{2l}(x_l,z_l))|x_l,z_l\rangle.
\end{align}
The vector  $\psi_{1a}\in\,\mathcal{H}_F$ describes the state that the observer $1$  associates to the particle $a$ by measuring statistical properties of observables; the vector $\psi_{2a}$ describes the  state that the observer $2$ can  associate to the particle $a$, etc... The spacetime coordinate $x$ is variable, indicating the spacetime localizations where the measurements can potentially be performed. Note that this state decomposition is not unique and depends on the choice of the observables at $1$ and $2$.
 Indeed, the observers $1$ and $2$ have the freedom to choose different observables to be measured. This freedom in the choice of the details of the experimental setting corresponds to different partitions of the sub-quantum degrees of freedom. From a quantum or classical description, it can be assumed  that the choice of the observable at $1$ is independent from the choice at $2$, but from the point of view of the sub-quantum description, a strong contextuality arises, which is manifest in a partition of the form
 \begin{align}
 \{(x_n,z_n)\}\mapsto \{(x_{1k},z_{2k}),(x_{1l},z_{2l})\}
 \label{first partition}
 \end{align}
The contextuality conditions  can change during the $\tau$-time, for instance, duration of the fly between the two particles. If the observers decide to change randomly the measurements settlements, this will imply a different partition
\begin{align}
 \{(x_n,z_n)\}\mapsto \{(x_{1'k},z_{1'k}),(x_{2'l},z_{2'l})\}.
 \label{second partition}
 \end{align}
 This can happen in a short time duration $\delta\tau$, but from the point of view of the $U_t$ dynamics, it happens in a long time duration of $t$-time. This provides a mechanism for interaction and correlation that can be different for each experimental setting. Moreover, this mechanism also shows a strong form of contextuality.
\begin{ejemplo}
When $a$ and $b$ are identical particles,  the states $\psi_{\lambda\beta}$ defined above correspond to one of the Bell states
\begin{align*}
\frac{1}{\sqrt{2}}(|00\rangle +|11\rangle ),\,\frac{1}{\sqrt{2}}(|01\rangle +|10\rangle ),\,\frac{1}{\sqrt{2}}(|01\rangle -|10\rangle )\,\textrm{or}\,\frac{1}{\sqrt{2}}(|00\rangle -|11\rangle ).
 \end{align*}
The appearance of a minus sign in some of these  linear combinations is associated to a factor $A_{\psi}$ containing a relative phase $\exp(\imath \pi)$ between the standard states associated to the quantum systems $a$ and $b$ relative to $1$ and to $2$. Note that for such relative phases to be local, that is, to be defined at the points $1$ and $2$, they must have the same value for all the sub-quantum degrees of freedom associated to $1$ respect to the degrees of freedom associated to $2$. Note that in doing this identification we are assuming that the spin degrees of freedom of quantum systems can be described by means of the poinwise sub-quantum degrees of freedom.
\end{ejemplo}

 An heuristic justification for the constraints \eqref{entanglementcondition2} can be introduced as follows. During the ergodic regime of the $t$-time evolution, the world lines of the sub-quantum degrees of freedom associated to the quantum systems $a$ and $b$ fill the full available phase space. If we assume that two generic sub-quantum degrees of freedom interact any time that there is a coincidence in the spacetime,
  \begin{align*}
  \vartheta_{k_a}(x_{ka})=\,\vartheta_{k_b}(x_{lb})\in\,M_4,
   \end{align*}
   then during the ergodic regime sub-quantum degrees associated to $a$ will interact with the sub-quantum degrees associated to $b$ in such a way that the thermalization conditions \eqref{entanglementcondition2} hold good.  It is also relevant to note that the formalism used is $4$-dimensional covariant. Therefore, the ergodicity of the evolution is embedded in the allowable domain of the four-dimensional spacetime $M_4$.  If the $t$-time world lines are continuous, this type of ergodicity and the application of a version of the ergodic theorem seems not to be  compatible with the existence of non-trivial causal relation, in particular, with the principle of maximal speed of light in vacuum. Thus this mechanism forces to consider $t$-time and the $U_t$ evolution as discrete. Furthermore, this mechanism may impose severe constraints on the interactions: if non-linear interactions terms are present, they can induce soliton solutions  after long time evolution and then the constraints \eqref{entanglementcondition2} could  not hold.

Let us consider the formal projection
\begin{align}
 pr:\mathbb{R}\times\mathbb{R}\to \mathbb{R},\quad (t,\tau)\mapsto \tau .
 \label{projection}
  \end{align}
This projection must appear implicitly in any effective mathematical description of Hamilton-Randers systems. It is a mathematical statement consistent with the interpretation of the $\tau$-time parameter as emergent. After the implementation of this projection in an effective mathematical description,  the details of the $U_t$ dynamics are not available any more. This means that the information about the sub-quantum filling of the available phase spacetime and interactions is unaccessible from the effective mathematical description. Usually, the effective mathematical description has as arena the four dimensional spacetime $M_4$, a four-dimensional manifold. Thus from such spacetime description, the correlations appear as instantaneous.

   The generalization of the mechanism to systems with more than two particles and more than two observers is straightforward. For instance, for  a system of three particles and three observers, the predecessor state is a product state of the form
     \begin{align*}
 &\Psi_{a \otimes b\otimes c}(u_a,u_b,u_c) =\,\Big(\sum^{N_a}_{k=1}\Big( n_{1k}(x_k,z_k)\,\exp(\imath \,\varphi_{1k}(x_k,z_k))\\
 & +n_{2k}(x_k,z_k)\,\exp(\imath \,\varphi_{2k}(x_k,z_k))+\,n_{3k}(x_k,z_k)\,\exp(\imath \,\varphi_{3k}(x_k,z_k))\Big)|\vartheta^{-1}_{k}(x),z_k\rangle\Big)\\
 & \otimes\Big(\sum^{N_b}_{l=1}\,\Big(n_{1l}(x_l,z_l)\,\exp(\imath \varphi_{1l}(x_l,z_l)\\
 & +n_{2l}(x_l,z_l)\,\exp(\imath \,\varphi_{2l}(x_l,z_l))+n_{3l}(x_l,z_l)\,\exp(\imath \,\vartheta_{3l}(x_l,z_l)) \Big)|\vartheta^{-1}_{l}(x),z_l\rangle\Big)\otimes\\
 &\Big(\sum^{N_q}_{q=1}\,\Big(n_{1q}(x_q,z_)\,\exp(\imath \varphi_{1q}(x_q,z_q)\\
 & +n_{2q}(x_q,z_q)\,\exp(\imath \,\varphi_{2q}(x_q,z_q))+\,n_{3q}(x_q,z_q)\,\exp(\imath \,\varphi_{3q}(x_q,z_q))\Big)|\vartheta^{-1}_{q}(x),z_q\rangle\Big).
 \end{align*}
 The corresponding quantum state is obtained by averaging along the speed coordinates and under the imposition of thermalization conditions on the $U_t$ dynamics corresponding to entanglement constraints.
The possibilities of entanglement for three particle systems is  richer  than for two particles, but the mechanism is analogous to the mechanism described above for two systems composed by two entangled particles and two observers.

 \section{General consequences of the theory} There are several general  consequences of our theory that are potentially testable against experiment. The first of these consequence is the following. In Hamilton-Randers theory, the sub-quantum degrees of freedom are constrained to be sub-luminal for the $U_t$ flow. This means that the rate of change coordinates  due to the $U_t$ dynamics respect to the arbitrary $t$-time parameter in a specific class of coordinate systems of $M$  is bounded by the speed of light in vacuum, when the change in position for light is parameterized using the same $t$-parameter. In order to understand why this must be the case, note that respect to any $\tau$-time parameter, the speed coordinate associated to the sub-quantum degrees of freedom is bounded by the speed of light. This is a consequence of the regularity of the underlying Randers-type structure of our models. If the $t$-velocity is not bounded by the speed of light, then it could happen that for particular configurations also the $\tau$-velocity is not bounded, in contradiction with the Randers conditions. However, if the formal projection $(t,\tau)\mapsto \tau$ is considered, there is a complete loss of details about the sub-quantum dynamics level. This implies that the interaction and correlation produced by the $U_t$ dynamics appear as  instantaneous in terms of the one-dimensional $\tau$-time parameter associated to the usual spacetime description of the dynamical system.

  In order to infer this consequence, an identification  between an  specific class of coordinate systems where Hamilton-Randers theory is formulated (\cite{Ricardo2017}, {\it section} 2 and {\it section 3}) and a class of coordinate systems associated to the  macroscopic observers that measure the apparent speed of the correlations has been done. Indeed, such macroscopic observers are identified with inertial observers or free falling coordinate systems in the most realistic situation when gravity is taken into account.

Given such coordinate systems, the surprise comes with the current experimental lower bounds for the apparent speed of the quantum correlations, bounds that surpass by several orders of magnitude the speed of light \cite{CocciaroFaettiFronzoni, GisinScaraniTittleZbinden, Salart et al., Juan Ying et al.}. This is in concordance with our mechanism for quantum entanglement in the context of Hamilton-Randers theory, that predicts  the apparent instantaneous macroscopic character for quantum correlations. This prediction is clearly falsifiable, since  it is enough to demonstrate experimentally the existence of an upper bound on the speed of quantum correlations measured in a inertial coordinate frame in a particular class of quantum entangled system to falsified our theory. Remarkably, our explanation of the apparent instantaneous speed of quantum correlations differs from purely operational treatment of quantum mechanics.

 The second consequence of the proposed mechanism for the quantum correlations, in this case originated by  the sub-luminal kinematics of the sub-quantum degrees of freedom and the filling mechanism of the $U_t$ flow, is the existence of a maximal spatial distance for the manifestation of the quantum correlations. Let us consider an inertial frame coordinate system and  assume that the source of the entangled system is located  at the origin of coordinates system. Then the causal spacetime domains for the sub-quantum degrees of freedom are bounded by the expression
\begin{align}
d_{cor}\leq \,c\,T.
\label{dcor}
 \end{align}
This relation is valid in any inertial coordinate system, associated with very specific local coordinate charts of  the configuration manifold $M$. By the relation \eqref{ETrelation}, the distance $d_{cor}$ depends upon the mass of the quantum system. In particular, one has the relation
 \begin{align}
d_{cor}\leq \,c \,T_{min}\,\exp\left[\frac{T_{min}\,M \,c^2}{\hbar}\right].
\label{boundfordistanceofcorrelations1}
 \end{align}
   For massless quantum particles, the equivalent expression  to \eqref{boundfordistanceofcorrelations1} is
  \begin{align}
d_{cor}\leq \,c \,T_{min}.
\label{boundfordistanceofcorrelations2}
 \end{align}
 If the system is composed by several non-identical particles, the semi-period $T$  in the relation \eqref{dcor} is the maximal of the individual semi-periods.

The actual ergodic motion and the use of the ergodic theorem in the form of the substitution of $t$-time average by phase average implies that the system passes many times by each point of the spacetime causal cone. Thus the distance $d_{corr}$ must be  much less than $T\,c$. Therefore, the above expressions \eqref{boundfordistanceofcorrelations1} and \eqref{boundfordistanceofcorrelations2} cannot not provide very accurate upper bounds for $d_{corr}$.

 The value of the universal semi-period $T_{min}$ is currently not specified by the theory. However, since it is known  the existence of quantum correlations for photons over macroscopic spatial distances, by the relation \eqref{boundfordistanceofcorrelations2}, $T_{min}$ must also be  macroscopic.  In this context, it is interesting to re-write \eqref{ETrelation} as
   \begin{align}
T=\,T_{min}\,\exp\left[\frac{T_{min}}{T_{P}}\,\frac{M}{M_{P}}\right],
\label{ET2}
 \end{align}
 where $T_{P}$ and $M_{P}$ are the time and Planck length. In this expression we explicitly observe the existence of two scales associated with the times $T_{min}$ and $T_{P}$ that play a relevant role in our theory. It is tempting to think $T_P$, the Planck time, as the fundamental step of time for the $U_t$ evolution. In such interpretation, the ${U}_t$ dynamics and the $t$-time parameters  must be discrete. Also, the condition 
 \begin{align*}
 T_{min}\gg T_{P}
 \end{align*}
 must hold, in order to apply the ergodic theory as before.

  The mechanism for quantum correlations described above implies that if the distance associated to the measurement positions at the location $1$ and at the location $2$ is larger than $d_{cor}$, then the correlation must cease. To illustrate how this happens, let us remark that the quantum correlation described by the constraints on each pair $(k,l)$ of the form \eqref{entanglementcondition2} is a consequence of an ergodic dynamics. If the two particles $a$ and $b$ defining the whole system  $a\cup b$ are separated by a short enough distance, the interacting mechanics persists and all the pairs $(k',l')$ will satisfy the relation \eqref{entanglementcondition2}. If the distance between $1$ and $2$ for the quantum correlation observers is increased long enough up to a distance comparable to $d_{cor}$, then none of the pairs $(k,l)$ satisfies the constraint   \eqref{entanglementcondition2}. Thus a transition
 \begin{align}
 \psi_{ab}\longrightarrow \psi_{a\otimes b}
 \label{transitionduringmeasurement}
 \end{align}
  must happen after the parts $a$ and $b$ of the system $a\cup b$ are far enough separate.
  $\psi_{a\otimes b}$ is a product state obtained after averaging along speed directions to the pre-quantum state,
 \begin{align*}
  \psi_{a \otimes b}(x) =\,&\Big(\sum^{N_a}_{k=1}\int_{T_uM^k_4}\,d^4z_k\,\Big( n_{1k}(x_k,z_k)\,\exp(\imath \,\vartheta_{1k}(x_k,z_k))\\
 & +n_{2k}(x_k,z_k)\,\exp(\imath \,\vartheta_{2k}(x_k,z_k))\Big)|(x_k,z_k\rangle\Big)\otimes\\
 &\Big(\sum^{N_b}_{l=1}\,\int_{T_uM^k_4}\,d^4z_l\,\Big(n_{1l}(x_l,z_l)\,\exp(\imath \vartheta_{1l}(x_l,z_l)\\
 & +n_{2l}(x_l,z_l)\,\exp(\imath \,\vartheta_{2l}(x_l,z_l)) \Big)|x_l,z_l\rangle\Big),
 \end{align*}
 or
 \begin{align*}
 \psi_{a \otimes b}(x)=\,\frac{1}{\sqrt{4}}\Big((\psi_{1a}(x)+\,\psi_{2a}(x))\otimes\,(\psi_{1b}(x)+\psi_{2b}(x))\Big).
 \end{align*}
 For instance, for each of the Bell states  $\frac{1}{\sqrt{2}}(|00\rangle +|11\rangle )$ and $\frac{1}{\sqrt{2}}(|01\rangle +|10\rangle )$ the final state is the same vector  $\psi_{a \otimes b}$ given by
 \begin{align*}
 \psi_{a\otimes b}=\,\frac{1}{\sqrt{4}}\Big((|0\rangle_a+\,|1\rangle_a)\otimes\,(|0\rangle_b+\,|1\rangle_b)\Big),
 \end{align*}
while for the Bell states $\frac{1}{\sqrt{2}}(|01\rangle -|10\rangle ),\frac{1}{\sqrt{2}}(|00\rangle -|11\rangle )$,
the final state $\psi_{a\otimes b}$ is of the form
 \begin{align*}
 \psi'_{a\otimes b}=\,\frac{1}{\sqrt{4}}\Big((|0\rangle_a-\,|1\rangle_a)\otimes\,(|0\rangle_b+\,|1\rangle_b)\Big).
 \end{align*}
 Thus,  the final states $\psi_{a\otimes b}$ are limit cycles where different entangled states evolve after  long $U_t$ evolution.

 The $\tau$-time rate of the transition \eqref{transitionduringmeasurement} depends on the details of the ergodic regime of the $U_t$ dynamics. It is reasonable to assume that the process is smooth and that the entanglement is lost slowly as a function of the spatial separation between $1$ and $2$.

 Note that the transition \eqref{transitionduringmeasurement} does not correspond to an spontaneous collapse as  usually considered in quantum mechanics (\cite{Von Neumann 1933 version 1955}, {\it Chapter VI}). In quantum mechanics, a measurement process implies the transition from a pure to a mixture state, according to the collapse postulate. In contrast, the final state $\psi_{a \otimes b}$ is not a mixture, but a pure quantum state and it is not associated with any particular quantum measurement.

   \section{Discussion: non-local character of the quantum correlations and the problem of signaling in Hamilton-Randers theory and some other remarks}
     Hamilton-Randers theory supports the point of view that quantum correlations have a non-local spacetime character. The theory is consistent with the violation of the Bell inequalities \cite{Bell1964,Bell} and also suggests a mechanism for the non-local character of quantum correlations  based upon the non-local and  partially ergodic property of the sub-quantum $U_t$ dynamics. Such non-locality avoids the main consequences of Bell's theory (see \cite{Ricardo2017}, {\it section 8}). The main argument given was that the expectation value of observables given by Hamilton-Randers theory for the case of entangled states (\cite{Ricardo2017}, {\it section 6}) do not coincide with the expectation value given by Bell's theory. In particular, Bell inequalities does not hold in Hamilton-Randers theory because the fundamental relation 
     \begin{align}
P(\vec{a},\vec{b})=\int \,d\lambda\,\rho(\lambda)\,A_1(\vec{a},\lambda)\,B_2(\vec{b},\lambda),
\label{bellprobabilitydistribution}
\end{align}
used in Bell's theory \cite{Bell1964,Bell} in the derivation of Bell inequalities, does not hold in Hamilton-Randers theory. Instead of this type of expression, in Hamilton-Randers theory a more complicated expression arises (\cite{Ricardo2017}, {\it section 8}).

      Moreover, the distribution of the  of sub-quantum degrees of freedom has a rather remarkable contextual character. This is directly observed from the partitions \eqref{first partition} and \eqref{second partition}, which depend on the choice of the observables to be measured at $1$ and at $2$. Therefore, it is not possible to write down a putative function determining the real values of any admissible real value with  all the functional properties required by Kochen-Specken theorem \cite{Kochen-Specken1967,Isham1995}. Thus, the mechanism discussed can potentially  avoid Kochen-Specken theorem and its consequences.

Although the $U_t$ dynamics is non-local from the point of view of spacetime, it is local from the point of view of the detailed description in the configuration space $TM$. Should we feel satisfied with our objective to understand quantum correlations' mechanism from a geometric point of view? For us, the explanation given by our theory is a geometric, objective explanation of quantum correlation. The relation with the spacetime description is through the histories transplanted from $TM$ to $TM_4$ by the diffeomorphisms induced from the collection $\{\vartheta_k\}$. The history described is non-local in $TM_4$, but it is clear, geometric and potentially  falsifiable. Also, note that the theory provides an explanation for the principles of diffeomorphism invariance, since it is required by consistency: it is not possible to describe in detail the geometry of the sub-quantum degrees of freedom, in particular it is not possible to specified the particular diffomorphisms $\{\vartheta_k\}$. Thus, all possible diffeomorphisms are equally valid and the effective theories must be constructed in a $\Diff (M_4)$-invariant way \cite{Ricardo2017}.

 According to Hamilton-Randers theory, a new notion of time, in particular for the notion of {\it time parameter}, must be adopted in the description of the dynamics of physical systems. It is not only that a notion of two-dimensional time substitutes the usual notion of one-dimensional time used in quantum mechanics and classical and quantum field theory, but also and motivated by our interpretation of quantum correlations between quantum systems separated by macroscopic distances, the domain of the  $t$-time parameters  used in the description of the $U_t$ dynamics of the fundamental cycles  must be of  macroscopic size, not only in the sense  that $T/T_P\gg 1$, but also that the correlations are over macroscopic distances. Although this fact does not imply the  enlargement of the spacetime metric structure from four to five dimensional metric structures, to have a consistent picture  one needs to address why macroscopic observers apparently do not notice such macroscopic $t$-time dimension. A possible way to solve this dilemma  is based on the observation that spacetime dimensions are probed by   classical and quantum test systems that interact through quantum gauge interactions or by means of the classical gravitational interaction. But Hamilton-Randers theory suggests that all the interactions of the Standard Model of particle physics and gravity are perceived in the domain where the Hamilton-Randers systems are naturally collapsed. The exception to this rule is the effects that macroscopic observers can perceive with direct origin on the mixing mechanism of the $U_t$ dynamics, namely, quantum correlations.  Therefore, the way in which  the second dimension of time  shows up is through the phenomenology related to quantum correlations, while any other phenomena is well described by one-time dynamical models.

 This new interpretation of time has as a significant consequence the emergence of the $\tau$-time, in particular, the natural irreversibility of the $U_\tau$ evolution. Despite that the physical laws associated to the $U_\tau$ evolution are reversible, the intrinsic mechanism that generates any $\tau$-time parameter implies that the corresponding intrinsic irreversible character of the reality.
 
 The idea that time could be two-dimensional is not new in theoretical physics. A comprehensive theory of {\it two-time physics}  was developed by I. Bars in the last decades in a context of string theory \cite{Bars2001}. However, Bars' notion of two-dimensional time is different from ours in several fundamental aspects. Probably the most relevant differences are first, the way in which $\tau$-time is emergent and the discrete character of $(t,\tau)$ in Hamilton-Randers theory, in contrast with the continuous character of time in Bars' theory. Second, the way the two-time is geometrized in Bars' theory, a geometrization which is absent in Hamilton-Randers theory.

 We have discussed several consequences of our theory. In particular, the limited range of the quantum correlations, showing its dependence with   the mass $M$ of the system as well as the occurrence of the generic process \eqref{transitionduringmeasurement}, are deviations from the predictions of standard quantum mechanics. The possibility to test these qualitative predictions is remarkable, since they are deviations from standard quantum mechanics.

 If the standard spontaneous collapse postulate of quantum mechanics is interpreted as representing a real physical process, a paradoxical issue that appears is the possibility of  signaling faster than the speed of light \cite{Gisin1991}.  Gisin's argument is based upon an application of Schmidt's decomposition  \cite{Schmidt1907} and the collapse postulate of quantum mechanics \cite{Von Neumann 1933 version 1955}. The argument shows that in principle mixtures evolving in different ways can be prepared at a distance by choosing conveniently the measurement arrangements of the separate observers and that this activates the possibility for faster than light signaling. Is  Gisin's argument applicable to Hamilton-Randers models? It seems that this is not the case. In our theory there is no spontaneous collapse processes as in quantum mechanics or objective collapse models. Instead, there are natural spontaneous collapse processes, that happen even in the absence of measurement. These processes are, and this a key point in our argument, {\it partially reversible} or at least,  are not permanent. According to our theory, every quantum mechanical system suffers a natural spontaneous collapse during the contractive regime in each fundamental cycle of the $U_t$ dynamics. Therefore, in the framework of Hamilton-Randers theory   is not possible to signaling faster than the speed of light by application of  the procedure described in \cite{Gisin1991}, since there is no final and persistent collapsed state.

 On the other hand, the mechanism for signaling discussed by Bancal et al. was based upon the assumption that quantum correlations propagate in spacetime faster than the speed of light  but  with finite speed respect to a particular class of preferred coordinate systems. However,  in Hamilton-Randers theory  the quantum correlations are apparently instantaneous from the point of view of the spacetime description of quantum systems. Thus, the argument for signalling discussed in \cite{Bancal et al} is avoided in our theory too. This is consistent with the fact that in Hamilton-Randers theory there is no a privileged reference frame respect to which sub-quantum degrees of freedom move faster than the speed of light \cite{Ricardo2017}.

 Our introduction of the partial ergodic regime, the association of the quantum wave function and the trace operation on the velocity coordinates can be consistent only if the evolution is discrete (\cite{Ricardo2017}, {\it section 5}). The current continuous formulation of the theory is an approximation.

    Finally, let us remark that the transitions \eqref{transitionduringmeasurement} and their generalizations to multiple particle states are deterministic, since the $U_t$ dynamics is  deterministic. It is also a local evolution in the configuration space $TM$. There are two possibilities to understand these processes. First, if the evolution of the quantum particles $a$ and $b$ is a free evolution, the transition must be driven by a different dynamics than  Heisenberg equation. The other possibility is that if the systems $a$ and $b$ are entangled, then they  cannot be considered free. The type of interactions that they suffer are  sub-quantum interactions but the slow dynamics is still of Heisenberg's type.

 \subsection*{Acknowledgements}  This work is supported by the Foundational Questions Institute (FQXi).

\end{document}